\RequirePackage{fix-cm}
\documentclass{SCIS2026}
\usepackage{amsmath}
\allowdisplaybreaks

\providecommand{\IEEEmembership}[1]{}
\newenvironment{IEEEbiography}[2][]{\begingroup\setbox0=\vbox\bgroup\noindent\textbf{#2}\par}{\egroup\endgroup}

\hypersetup{hidelinks}

\def\ban{\begin{eqnarray*}}
\def\ean{\end{eqnarray*}}
\def\bna{\begin{eqnarray}}
\def\ena{\end{eqnarray}}

\begin{document}	

 \title{Tracking analysis of compressed-communication distributed adaptive estimation for time-varying parameters}{Tracking analysis of compressed-communication distributed adaptive estimation for time-varying parameters}

\ArticleType{RESEARCH PAPER}
\Year{2026}
\Month{}
\Vol{}
\No{}
\DOI{}
\ArtNo{}
\ReceiveDate{}
\ReviseDate{}
\AcceptDate{}
\OnlineDate{}
\AuthorMark{Gan D, et al}
\AuthorCitation{Gan D, Chen S N, Wang Y}

\author[1,2]{Die GAN}{}
\author[3]{Shuning CHEN}{}
\author[4,3]{Ying WANG}{wangying96@amss.ac.cn}

\address[1]{School of Mathematical Sciences, Fudan University, Shanghai 200433, China} \address[2]{Laboratory of Mathematics for Nonlinear Science, Fudan University, Shanghai 200433, China}
\address[3]{State Key Laboratory of Mathematical Sciences, Academy of Mathematics and Systems Science,\\ Chinese Academy of Sciences, Beijing 100190, China}
\address[4]{Department of Decision and Control Systems, Royal Institute of Technology, Stockholm 11428, Sweden}

\abstract{
This paper studies the problem of distributed time-varying parameter tracking over communication-constrained networks.
To mitigate the accumulation of induced compression errors across the network, we propose a differentially compressed distributed least mean squares (LMS) algorithm. 
In the proposed scheme, each sensor updates a local auxiliary state recursively based on the compressed difference between the current parameter estimate and the previous auxiliary state, and then transmits only this compressed difference to its neighbors for auxiliary-state reconstruction. 
This differential recursion mechanism helps control the accumulation of compression errors and accommodates a general class of randomized compressors including randomized quantizers and sparsifiers. 
We establish the mean-square boundedness  and explicit steady-state tracking-error bound under a cooperative excitation condition on stochastic regression vectors, which captures both temporal and spatial information to reflect the joint excitation effect of sensors. 
Notably, our stability analysis requires neither temporal independence nor stationarity of regressors, nor persistent excitation at individual nodes,  thereby accommodating correlated and nonstationary signals  from stochastic feedback systems.
Numerical simulations demonstrate  the theoretical results under correlated and non-stationary regression data.
}


\maketitle

\section{Introduction}
Distributed adaptive filtering and estimation have been widely applied in various fields, including cyber-physical systems, environmental monitoring, target tracking, and networked control \cite{Sayed2014,Dimakis2010,RSCB2014}. In these applications, spatially distributed sensors collaboratively estimate unknown and possibly time-varying parameters using local measurements and information exchanged with neighboring sensors \cite{Gan2021Stability}. Compared with centralized processing, such distributed architectures avoid
transmitting all raw measurements to a fusion center and therefore offer
improved scalability and robustness in large-scale networks. Motivated by these advantages and the growing demand for efficient information fusion and collaborative decision-making, recent studies have further advanced distributed estimation and cooperative learning in multi-agent systems \cite{Dai2026Dynamic,Zhang2023Distributed}.

Among distributed adaptive estimation methods, distributed least mean squares (LMS) algorithms are particularly attractive for real-time estimation and tracking because of their simple recursive structure and low computational complexity \cite{LopesSayed2008,CattivelliSayed2010}. 
Classical analyses of diffusion-beased and consensus-based distributed LMS algorithms are developed under statistical assumptions such as stationary regression processes and temporal or spatial independence \cite{CattivelliSayed2010,MateosEtAl2009,ZhaoSayed2012,ChenSayed2012}. 
Although these assumptions facilitate performance analysis, they are restrictive for regressors generated by feedback systems, stochastic processes, or correlated sensing environments. 
Building on the conditional-expectation-based excitation condition introduced in \cite{Guo1990Estimating}, Xie and Guo \cite{XIE2018,XieGuo2018} established the stability theory for distributed  LMS-based  adaptive filters under  a cooperative excitation condition. 
These results provide important stability guarantees for distributed adaptive estimation under general data conditions. However, they rely on frequent exchange of full-precision, uncompressed information among neighboring sensors, both of which are difficult to satisfy in practical networks subject to bandwidth constraints.

On the communication side, transmitting full parameter vectors at every iteration in classical distributed LMS \cite{CattivelliSayed2010,MateosEtAl2009,ZhaoSayed2012,ChenSayed2012,XIE2018,XieGuo2018}  incurs prohibitive costs in high-dimensional or bandwidth-constrained networks.
To reduce overhead, quantization and sparsification have been widely investigated for communication-constrained distributed algorithms \cite{Cao2023}. Existing studies can be roughly categorized into three directions: direct compression of iterates \cite{YuanEtAl2012,Alistarh-2017,StichEtAl2018,WangniEtAl2018}, reconstruction-based encoding \cite{SayinKozat2014, Wang2026Distributed}, and differential compression with auxiliary states \cite{pmlr-v130-kovalev21a,RichtarikEtAl2021,CondatEtAl2022}. 
In particular, randomized compression operators satisfying unbiasedness and bounded relative mean-square error provide a unified framework for randomized quantization and sparsification \cite{KoloskovaEtAl2019}.
However, directly compressing current iterates causes compression errors to re-enter the recursion repeatedly, leading to information loss and error accumulation \cite{StichEtAl2018,KoloskovaEtAl2019}. 
While auxiliary states and differential compression effectively mitigate error accumulation in distributed optimization, their convergence theories are primarily developed for time-invariant objective functions. 
In contrast to optimization problems with time-invariant objectives, tracking stochastic time-varying parameters driven by system dynamics introduces persistent parameter variations into the recursion, which causes compression errors to interact with parameter drift and stochastic regressors.
This fundamental difference prevents direct extension of existing compression theories from optimization to adaptive tracking.

Efforts have also been made to incorporate communication compression into distributed adaptive estimation. 
Sayin and Kozat \cite{SayinKozat2014} proposed projection-based compressive diffusion strategies that reduce local parameter estimates to scalar or single-bit messages, with estimation performance analyzed under Gaussian regression and independence assumptions. 
Wang et al. \cite{Wang2026Distributed} later studied the related reconstruction-based ideas in quantized distributed identification, where a persistently exciting compression encoding scheme was employed to estimate a time-invariant parameter from one-bit communication.
These schemes reconstruct neighbors’ parameter estimates from compressed messages for use in local estimation updates.
Harrane et al. \cite{HarraneEtAl2019} developed a direct-compression diffusion LMS strategy, in which selected entries of local estimates and gradient vectors are directly transmitted, again for time-invariant parameter estimation. 
Carpentiero et al. \cite{Carpentiero-2024} studied distributed compressed adaptive learning  under independence assumptions on regression data. 
More recently, Yuan et al. \cite{Yuan2026} introduced trainable quantization with error compensation into diffusion LMS, but their analysis also targets time-invariant parameters under standard independence assumptions.

Despite these advances, existing compressed distributed LMS methods mainly address time-invariant parameter estimation or learning \cite{Wang2026Distributed,HarraneEtAl2019,Carpentiero-2024,Yuan2026}, or rely on specialized compression structures and restrictive statistical assumptions on the regressors \cite{SayinKozat2014,HarraneEtAl2019,Carpentiero-2024,Yuan2026}.
To the best of our knowledge, a stability theory for compressed distributed LMS
tracking  under general data conditions is few.
The main challenge is that compression errors are recursively fed back into the adaptive estimation process and interact with parameter drift and stochastic regressors, which
 calls for a new stability analysis for compressed distributed LMS tracking under general stochastic regressors.

Motivated by the above considerations, this paper studies distributed LMS tracking of stochastic time-varying parameters under communication constraints, with a focus on compression-error control and stability analysis. We develop a differentially compressed distributed LMS algorithm for this tracking problem. The proposed algorithm integrates differential auxiliary-state updates with normalized local innovations and neighbor-consensus corrections. Specifically, each sensor recursively updates its auxiliary state using a randomized compressed difference between its current estimate and the previous auxiliary state. This compressed difference is then transmitted to neighboring sensors for auxiliary-state reconstruction, thereby reducing communication overhead. The updated auxiliary states are used to compute local observation residuals and construct consensus corrections, coupling local adaptation with inter-sensor cooperation. Note that the proposed scheme accommodates a class of unbiased randomized compressors, including randomized quantizers and sparsifiers.
For theoretical analysis, we establish mean-square boundedness under a cooperative excitation condition on the stochastic regressors. We also derive an explicit bound on the steady-state tracking error. To handle the recursive coupling between compression and estimation errors, we derive a mean-square contraction inequality by jointly analyzing the auxiliary-state and parameter-estimation errors over multiple steps. Unlike \cite{SayinKozat2014,HarraneEtAl2019,Carpentiero-2024,Yuan2026}, our analysis requires neither temporal independence nor stationarity of the regressors, nor persistent excitation at individual sensors, which accommodates correlated and non-stationary signals generated by stochastic feedback systems.

The remainder of this paper is organized as follows. Section \ref{problem_formulation} presents the problem formulation. Section \ref{Algorithm design} introduces the randomized compression operators and the proposed compressed-communication distributed LMS algorithm. 
Section \ref{main_results} gives the main results for the proposed algorithm. 
The proofs are provided in Section \ref{proofs_of}. 
Section \ref{simulaiton} presents numerical simulations, and Section \ref{concluding} concludes the paper with a discussion of future research directions.

\section{Problem formulation}\label{problem_formulation}

\subsection{Basic notations}
Throughout this paper, for a matrix $X \in \mathbb{R}^{m \times n}$, we denote $\|X\| = \left( \lambda_{\max}\left(X X^\top\right) \right)^{\frac{1}{2}}$, where $^\top$ denotes the matrix transpose, and $\lambda_{\max}(\cdot)$ and $\lambda_{\min}(\cdot)$ denote the largest and smallest eigenvalues, respectively. We use $I$ to denote the identity matrix with suitable dimensions.
Moreover, $\mathbb{E}[\cdot]$ and $\mathbb{E}[\cdot|\cdot]$ represent the expectation and conditional expectation, respectively.
The operators $\mathrm{col}(\cdot,\cdots,\cdot)$ and $\mathrm{diag}(\cdot,\cdots,\cdot)$ denote vector stacking and block-diagonal construction, respectively.

\subsection{Communication topology}
A fixed weighted undirected graph $\mathcal{G}=(\mathcal{V},\mathcal{E},\mathcal{A})$ describes the communication topology, where $\mathcal{V}=\{1,2,\cdots,n\}$ is the sensor set, $\mathcal{E}\subseteq\mathcal{V}\times\mathcal{V}$ is the edge set, and $\mathcal{A}=[a_{ij}]$ is a symmetric stochastic adjacency matrix.
The weights satisfy $a_{ij}=a_{ji}\geq0$ and $\sum_{j=1}^{n}a_{ij}=1$; for $i\ne j$, $a_{ij}>0$ if and only if sensors $i$ and $j$ are connected by an edge.
The neighbor set of sensor $i$ is $\mathcal{N}_i=\{j\ne i:a_{ji}>0\}\cup\{i\}$, including the sensor itself.
The graph is connected if every pair of distinct sensors is linked by a path.
The graph Laplacian is $\mathcal{L}=I-\mathcal{A}$, which is symmetric and positive semidefinite. If $\mathcal{G}$ is connected, then $\mathcal{L}$ has a simple zero eigenvalue and all its remaining eigenvalues are positive.

\subsection{Observation model}
Consider the time-varying parameter tracking problem over a network of $n$ sensors.
The data $\{y_{k,i}, x_{k,i}, k=0,1,2,\cdots\}$ collected by the sensor $i$ follows the discrete-time stochastic regression model,
\begin{equation}\label{e:xt}
	y_{k,i}=x_{k,i}^\top\xi_k+d_{k,i}, ~k=0, 1, 2, \cdots,
\end{equation}
where $y_{k,i}$ is the scalar observation or output of the sensor $i$ at time $k$, $x_{k,i}\in \mathbb{R}^m$ is the
stochastic regression vector,  $\xi_k\in \mathbb{R}^m$ is an unknown time-varying $m$-dimensional parameter, and $d_{k,i}$ is the observation noise. 
We further denote the variation of the parameter vector $\xi_k$ by
\begin{equation}\label{div}
\gamma\omega_k\triangleq\xi_k-\xi_{k-1},~k\ge1,
\end{equation}
where $\gamma$ denotes the parameter variation rate, and $\omega_k$ is an $m$-dimensional random variable. The time-invariant case is included as a special case with  $\gamma=0$.

The classical consensus-based distributed LMS algorithm (cf., \cite{XIE2018}) performs collaborative parameter estimation through information exchange among neighboring nodes. Its update is given by:
\begin{align}
	\hat{\xi}_{k+1,i} = \hat{\xi}_{k,i}+\mu\left\{\frac{x_{k,i}}{1+\|x_{k,i}\|^2}(y_{k,i}-x_{k,i}^\top\hat{\xi}_{k,i})-\nu\sum_{l\in\mathcal{N}_i}a_{li}(\hat{\xi}_{k,i}-\hat{\xi}_{k,l})  \right\},\label{NLMS}
\end{align}
where $\mu$ is the adapt rate, and $\nu$ is the weight coefficient of the consensus term.

However, real-time tracking using algorithm (\ref{NLMS}) requires frequent information exchange, resulting in substantial communication overhead. 
Communication compression can reduce this burden, but it also introduces additional errors into the adaptive recursion and complicates the stability analysis. Therefore, our objective is to develop a compressed-communication distributed adaptive filter for tracking \(\xi_k\)  and to establish its stability under general stochastic regressors,
without requiring independence or stationarity assumptions on $\{x_{k,i}\}$.

\section{\!Compressed-communication distributed  LMS algorithm}\label{Algorithm design}
In the following, we first introduce  a class of random compression operators with bounded relative compression error, based on which we develop a novel distributed filtering algorithm.

\subsection{Compression operators}
We consider randomized compression operators that preserve the input in expectation and have bounded relative mean-square compression errors.
\vspace{-5pt}
\begin{assumption}\label{assum:1}
	Given an input \( x \in \mathbb{R}^m \), there exists a minimal constant \( \rho > 0 \) such that the random compression operator \( {Q}:\mathbb{R}^m \mapsto \mathbb{R}^m \) satisfies 
	\begin{align}
	&\mathbb{E}\left[{Q}(x)-x \right]=0,\notag\\
	&\mathbb{E}\left\|{Q}(x)-x \right\|^2\le\rho\|x\|^2,\notag
	\end{align}
	where \( \rho \) is termed the ``compression parameter" of the operator. 	Additionally, the random compressed value \( Q(x) \) depends solely on the input \( x \)  and is mutually independent across individual instances.
\end{assumption}

\begin{remark}\label{re1}
	Note that the expectation in Assumption \ref{assum:1} is taken with respect to the randomness of the compression operator (i.e., for a deterministic \( x \)). If the input \( x \) is random, the corresponding conditional expectation properties hold:
	\begin{align}
		&\mathbb{E}\left[{Q}(x)-x|x \right]=0,\label{co:3}\\
		&\mathbb{E}\left[\left\|{Q}(x)-x \right\|^2\big|x\right]\le\rho\|x\|^2.\label{co:4}
	\end{align}
From Equation (\ref{co:4}), it can be observed that a smaller compression parameter $\rho$ corresponds to lower data distortion. This indicates that the data  undergoes only mild compression and can better preserve the characteristics of the original data. Conversely, a larger $\rho$ value implies that the data has been subjected to coarser quantization processing. While this enables a higher compression ratio, it introduces more significant information loss. This reveals the critical role of the compression parameter $\rho$ in balancing the trade-off between the degree of data compression and the preservation of data quality.
	\end{remark}

Many commonly used compression operators satisfy Assumption \ref{assum:1}, including randomized quantization operators and randomized sparsification operators \cite{Alistarh-2017, pmlr-v130-kovalev21a}.
We next briefly introduce these two representative classes. Randomized quantization reduces communication by lowering the representation precision of all components, whereas randomized sparsification reduces communication by transmitting only a randomly selected subset of components.

\begin{example}[Randomized quantization operators]
\label{RQO}
Given $x\in\mathbb{R}^m$, the randomized quantization operator represents each component using its sign and a randomized quantization of the normalized magnitude
\[
    \chi_j=\frac{|x_j|}{\|x\|}\in[0,1],\qquad j=1,\ldots,m,
\]
while transmitting the norm $\|x\|$ with $h$ bits. Let $Q_r(0)=0$. For $x\neq0$, partition $[0,1]$ into $L=2^r-1$ equal subintervals, where $r$ is the number of bits for each normalized magnitude. With
$ f(\chi_j)=\lfloor L\chi_j\rfloor,$
define the randomized rounding
\[
    f_r(\chi_j)=
    \begin{cases}
    f(\chi_j)+1, & \text{with probability }L\chi_j-f(\chi_j),\\
    f(\chi_j), & \text{otherwise}.
    \end{cases}
\]
The compressed vector is then given by
\[
    [Q_r(x)]_j
    =
    \|x\|\operatorname{sgn}(x_j)\frac{f_r(\chi_j)}{L},
    \qquad j=1,\ldots,m .
\]
The total communication cost is $h+m(r+1)$ bits, including one global norm, one sign bit, and $r$ magnitude bits per component, compared with $mh$ bits for full-precision transmission. Hence, when $r+1<h$, communication is reduced, and for $m\gg1$ the compression ratio approaches $
    \frac{mh}{h+m(r+1)}
    \approx \frac{h}{r+1}.$
The operator satisfies Assumption~\ref{assum:1}. The randomized rounding is unbiased, so $\mathbb{E}[Q_r(x)-x]=0 $. As shown in \cite{Alistarh-2017},
\[
    \mathbb{E}\|Q_r(x)-x\|^2
    \leq \rho\|x\|^2,
    \quad
    \rho=
    \min\left\{
    \frac{m}{(2^r-1)^2},
    \frac{\sqrt{m}}{2^r-1}
    \right\}.
\]
Thus, increasing $r$ improves compression accuracy at the cost of increased communication.
\end{example}

\begin{example}[Randomized sparsification operators]
Given $x\in\mathbb{R}^m$, let $F_s(x)$ be obtained by uniformly selecting $s<m$ components of $x$ and setting the remaining components to zero. The randomized sparsification operator is
\[
    Q_s(x)=\frac{m}{s}F_s(x),
\]
where the factor $m/s$ compensates for the random removal of components. Since each component is selected with probability $s/m$, $\mathbb{E}[Q_s(x)]=x$.
Only the $s$ selected components and their indices are transmitted. If each value requires $h$ bits and each index requires approximately $\log_2(m)$ bits, the communication cost is $ s\big(h+\log_2(m)\big).$
Thus, compared with $mh$ bits for full-vector transmission, substantial communication savings are obtained when $s\ll m$.

The operator also satisfies Assumption~\ref{assum:1}, since
$\mathbb{E}[Q_s(x)-x]=0$ and, as shown in \cite{pmlr-v130-kovalev21a},
\[
    \mathbb{E}\|Q_s(x)-x\|^2
    \leq \rho\|x\|^2,
    \qquad
    \rho=\frac{m}{s}-1.
\]
Hence, $s$ controls the trade-off between communication efficiency and compression accuracy: $s=m$ yields the identity mapping with $\rho=0$, whereas smaller $s$ reduces communication at the cost of larger compression error.
\end{example}

\subsection{Algorithm design}
Based on the above analysis of the properties of randomized compression operators, it can be observed that while reducing communication costs, such operators can introduce compression errors in a controllable manner. 
We next incorporate such operators into distributed LMS tracking. 

An intuitive approach is to directly combine the compression operator with the conventional LMS algorithm (\ref{NLMS}), replacing the original parameter estimate $\hat{\xi}_{k,i}$ with its compressed version $Q(\hat{\xi}_{k,i})$, thereby leading to the following update strategy:
\begin{align*}
\hat{\xi}_{k+1,i} = Q(\hat{\xi}_{k,i})+\mu\left[\frac{x_{k,i}}{1+\|x_{k,i}\|^2}\left(y_{k,i}-x_{k,i}^\top Q(\hat{\xi}_{k,i})\right)-\nu\sum_{l\in\mathcal{N}_i}a_{li}\left(Q(\hat{\xi}_{k,i})-Q(\hat{\xi}_{k,l})\right)  \right].
\end{align*}
However, directly compressing the complete estimates causes compression errors to repeatedly enter the adaptive recursion. 
These errors may accumulate over time and degrade convergence performance, or even lead to instability \cite{pmlr-v119-lu20a,JMLR:v24:21-1548}.
This motivates the use of an auxiliary compression state and differential compression, where only the difference between the current estimate and the auxiliary state is compressed and transmitted.

To mitigate the effects of compression errors, we introduce an  auxiliary state \( q_{k,i} \) and develop a differentially compressed-communication distributed LMS algorithm (CC-DLMS); see Algorithm \ref{algo:1}. The algorithm consists of two main steps:

\begin{itemize}
    \item 	{\textit{Step 1: Differential compression.}
    This step consists of transmission, reception, and decoding, as below:
    \begin{itemize}
        \item[--] \emph{Transmit:} Each sensor $i$ computes the difference vector $\delta_{k,i}\triangleq\hat{\xi}_{k,i}-q_{k,i}$ between its current estimate $\hat{\xi}_{k,i}$ and its auxiliary state $q_{k,i}$, compresses it using a random compression operator $Q(\cdot)$ satisfying Assumption \ref{assum:1}, and transmits $Q(\delta_{k,i})$ to its neighbors.
        \item[--] \emph{Receive:} Sensor $i$ receives compressed difference $Q(\delta_{k,l})$ from each neighbor $l\in\mathcal{N}_i\setminus\{i\}$.
        \item[--] \emph{Decode:} For every $l\in\mathcal{N}_i$, including $i$ itself, sensor $i$ updates the auxiliary state $q_{k,l}$ by adding the scaled compressed difference $\zeta Q(\delta_{k,l})$, as in (\ref{sys:compre}), where $\zeta\in(0,1)$ is a compression weight whose effect on stability is discussed in Section \ref{main_results}. The update for $l=i$ uses the same realization of $Q(\delta_{k,i})$ that was transmitted. With identical zero initialization and reliable message delivery before the estimate update, the locally stored copies coincide with the corresponding neighbors' auxiliary states.
    \end{itemize}}

	\item {\textit{Step 2: Estimate update.}
    Sensor $i$ then updates its estimate according to (\ref{sys:adapt}), using its own updated auxiliary state $q_{k+1,i}$ to evaluate the local innovation and the reconstructed neighbor states $q_{k+1,l}$, $l\in\mathcal{N}_i$, to form the consensus correction.}
\end{itemize}

The key advantage of this design lies in the fact that sensors only need to transmit the compressed difference vectors \(Q(\delta_{k,l})\), rather than exchanging complete parameter estimates.  Since the differences are formed relative to recursively updated auxiliary states, the transmitted information mainly reflects the mismatch between the current estimates and their auxiliary representations.
Meanwhile, the auxiliary recursion retains past information and helps compensate for compression errors.
As a result, the proposed design reduces communication overhead while controlling the accumulation of compression errors in the adaptive estimation process.

\begin{algorithm}[!ht]
	\caption{Compressed-communication distributed  LMS algorithm}\label{algo:1}
	\begin{algorithmic}
		\STATE {\bfseries Input data:} $\{y_{k,i},x_{k,i}\}_{i=1}^n$, $k=0,1,\cdots$
		\STATE {\bfseries Output:} $\{\hat{\xi}_{k+1,i}\}_{i=1}^n$, $k=0,1,\cdots$
		\STATE \textbf{Initialization:} For each sensor $i(i=1,\cdots,n)$, assign arbitrary initial estimates ${\hat{\xi}}_{0,i}$ {and initialize its own auxiliary state $q_{0,i}\in\mathbb{R}^m$ and the locally stored copies of its neighbors' auxiliary states to zero.}
		\STATE{At each time instant $k\in\mathbb{N}$:}
		\STATE{\quad For each node $i\in\{1,\cdots,n\}$: }
		\STATE {\qquad {\textbf{Step 1. Differential compression}}}
		\STATE {\qquad\quad - Transmit the compressed difference}
		\[
		{Q(\hat{\xi}_{k,i}-q_{k,i})}
		\]
		\STATE {\qquad\quad - Receive $Q(\hat{\xi}_{k,l}-q_{k,l})$ from $l\in\mathcal{N}_i\setminus\{i\}$}
		\STATE {\qquad\quad - Decode by computing}
		\begin{equation}\label{sys:compre}
		q_{k+1,l} = q_{k,l}+\zeta Q(\hat{\xi}_{k,l}-q_{k,l}),\quad \forall\, l\in\mathcal{N}_i
		\end{equation}
		\STATE {\qquad {\textbf{Step 2. Estimate update}}}
		\begin{align}
		\hat{\xi}_{k+1,i} = q_{k+1,i}+\mu\left[\frac{x_{k,i}}{1+\|x_{k,i}\|^2}\left(y_{k,i}-x_{k,i}^\top {q}_{k+1,i}\right)-\nu\sum_{l\in\mathcal{N}_i}a_{li}\left(q_{k+1,i}-q_{k+1,l}\right)  \right],\label{sys:adapt}
		\end{align}
		where $\mu$ is the adapt rate, and $\nu$ is the weight coefficient of the consensus term.
	\end{algorithmic}
\end{algorithm}

	\subsection{Recursive tracking error}
	To further analyze the tracking performance of the proposed algorithm, we first derive the corresponding tracking-error recursion. Define the parameter tracking error and auxiliary-state error as:
	\begin{align}
	\widetilde{\xi}_{k,i} \triangleq  \hat{\xi}_{k,i}-\xi_k
	\in\mathbb{R}^m,\quad
	\widetilde{ q}_{k,i} \triangleq   q_{k+1,i}-\xi_{k}\in\mathbb{R}^m.\label{q}
	\end{align}
	{Then, the difference variable satisfies}
	\begin{gather}
	\delta_{k,i} \triangleq \hat{\xi}_{k,i}-q_{k,i} =\tilde{\xi}_{k,i} -\tilde{q}_{k-1,i}+\gamma\omega_k\in\mathbb{R}^m,\label{e1}
	\end{gather}
	where $\omega_k$ is the parameter variation in  (\ref{div}), and $\gamma$ denotes the parameter variation rate.
	
	Combining the system equations (\ref{e:xt}) and (\ref{div}), the recursion (\ref{sys:compre}) and (\ref{sys:adapt}) in Algorithm \ref{algo:1} can be transformed into:
	\begin{align}
	\widetilde{{q}}_{k,i} &= \widetilde{{q}}_{k-1,i}+\zeta{Q}\left(\delta_{k,i}\right)-\gamma\omega_{k},\nonumber\\
	\delta_{k+1,i} &= \hat{\xi}_{k+1,i}-q_{k+1,i}\nonumber\\
	&=  \mu\left[\frac{x_{k,i}}{1+\|x_{k,i}\|^2}\left(y_{k,i}-x_{k,i}^\top {q}_{k+1,i}\right)-\nu\sum_{l\in\mathcal{N}_i}a_{li}\left(q_{k+1,i}-q_{k+1,l}\right)  \right]\nonumber\\
	&=  \mu\left[\frac{x_{k,i}}{1+\|x_{k,i}\|^2}\left(x_{k,i}^\top\xi_k-x_{k,i}^\top {q}_{k+1,i}+d_{k,i}\right)-\nu\sum_{l\in\mathcal{N}_i}a_{li}\left(\tilde{q}_{k,i}-\tilde{q}_{k,l}\right)  \right]\nonumber\\
	&=  -\mu\frac{x_{k,i}x_{k,i}^\top}{1+\|x_{k,i}\|^2}\tilde{q}_{k,i}+\mu\frac{x_{k,i}}{1+\|x_{k,i}\|^2}d_{k,i}-\mu\nu \sum_{l\in\mathcal{N}_i}a_{li}\left(\tilde{q}_{k,i}-\tilde{q}_{k,l}\right).\label{recur:1}
	\end{align}
	To facilitate the analysis of estimation performance for all sensors in the network, we introduce the following notations:
	\begin{align}
	&\delta_k \triangleq \text{col}\left\{\delta_{k,1},\cdots,\delta_{k,n}\right\}\in\mathbb{R}^{mn},\hskip 3.8cm
	\widetilde{ q}_k\triangleq \text{col}\left\{\widetilde{ q}_{k,1},\cdots,\widetilde{ q}_{k,n}\right\}\in\mathbb{R}^{mn},\nonumber\\
&X_k \triangleq \text{diag}\left\{ x_{k,1},\cdots, x_{k,n}\right\}\in\mathbb{R}^{mn\times n},\hskip 3cm
	\Omega_k\triangleq 1_n\otimes\omega_k\in\mathbb{R}^{mn},\nonumber\\
	& L_k \triangleq \text{diag}\left\{\frac{x_{k,1}}{1+\|x_{k,1}\|^2},\cdots,\frac{x_{k,n}}{1+\|x_{k,n}\|^2}\right\}\in\mathbb{R}^{mn\times n},\hskip 0.5cm
	F_k\! =\! L_k X_k^\top\!\in\!\mathbb{R}^{mn\times mn},\nonumber\\
&\mathscr{L}\!\triangleq\!\mathcal{L}\otimes I_m\!\in\!\mathbb{R}^{mn\times mn},\hskip 5.2cm
	\mathscr{A} \triangleq \mathcal{A}\otimes I_m\in\mathbb{R}^{mn\times mn},\nonumber\\
    &	\mathcal{ Q}(\delta_k)\triangleq \text{col}\left\{Q(\delta_{k,1}),\cdots,Q(\delta_{k,n})\right\}\in\mathbb{R}^{mn},\hskip 2.3cm
    G_k = F_k+\nu\mathscr{L}^{\top}\in\mathbb{R}^{mn\times mn},\nonumber\\
	&D_k = \text{col}\left\{d_{k,1},\cdots,d_{k,n} \right\}\in\mathbb{R}^{n},\hskip 4cm
	\widetilde{\Xi}_k \triangleq\text{col}\left\{\widetilde{ \xi}_{k,1},\cdots,\widetilde{\xi}_{k,n}\right\}\in\mathbb{R}^{mn}.\label{z}
	\end{align}
	Thus by (\ref{recur:1}) and (\ref{z}), the global recursive relation can be obtained:
	\begin{align}
	\delta_{k+1} &= -\mu F_k\widetilde{ q}_k+\mu L_kD_k-\mu\nu(I-\mathscr{A}^\top)\widetilde{ q}_k = -\mu G_k \widetilde{ q}_k+\mu L_kD_k,\label{e2}\\
	\widetilde{ q}_{k+1} &= \widetilde{ q}_k+\zeta\mathcal{ Q}( \delta_{k+1})-\gamma\Omega_{k+1} \nonumber= \widetilde{ q}_k+\zeta\delta_{k+1}+\zeta e_{k+1}-\gamma\Omega_{k+1}\nonumber\\
	&= \!(I\!-\!\zeta\mu G_k)\widetilde{ q}_k\!+\!\zeta\mu L_kD_k\!+\!\zeta e_{k+1}\!-\!\gamma\Omega_{k+1},\label{equ:2}
	\end{align}
	where $ e_{k+1} \triangleq \mathcal{ Q}( \delta_{k+1})- \delta_{k+1}$.

	Given that \(\widetilde{q}_k\) exhibits superior recursive properties compared to \(\widetilde{\xi}_k\) in this compression algorithm, the stability analysis of \(\widetilde{q}_k\) will be prioritized in the main results of the next section. By (\ref{e1}) and (\ref{e2}), the stability  for the tracking error \(\widetilde{\xi}_k\) will be further established through the following relation:
	\begin{align}\label{q-xi}
	\widetilde{\Xi}_{k+1} &= \widetilde{q}_{k} + \delta_{k+1} - \gamma\Omega_{k+1}= (I - \mu G_k)\widetilde{q}_{k} + \mu L_kD_k - \gamma\Omega_{k+1}.
	\end{align}

	\section{Stability Analysis of CC-DLMS Algorithm}
    \label{main_results}
	
	Before conducting the stability analysis, we first introduce some key assumptions and lemmas.
	\begin{assumption}\label{assum:g}
	Graph $\mathcal{G}$ is undirected and connected.
\end{assumption}

\begin{remark}\label{rmk:2}
	The eigenvalues of the Laplacian matrix \(\mathcal{L} = I - \mathcal{A}\) can be arranged in non-decreasing order as \(0 = \lambda_1(\mathcal{L}) \le \lambda_2(\mathcal{L}) \le \cdots \le \lambda_n(\mathcal{L}) \le 2\). Under Assumption \ref{assum:g}, it holds that \(\lambda_2(\mathcal{L}) > 0\) \cite{Godsil2001AlgebraicGT}.
\end{remark}

\begin{lemma}\cite{XIE2018}\label{lem:e}
	Under Assumption \ref{assum:g}, the matrix $G_k$ defined in (\ref{z}) satisfies the inequality $0\le \mu G_k\le I$ for all admissible parameters $\mu,\nu$ that fulfill the constraint condition $\mu(1+2\nu)\le1$.
\end{lemma}

\begin{lemma}\cite{XIE2018}\label{lem:sigma}
	Consider a sequence of positive semi-definite bounded matrices ${A_{k,i}\in\mathbb{R}^{m\times m},k\ge0 }(i = 1,\cdots,n)$ satisfying that there exists a constant $c$ such that $0\le A_{k,i}\le cI_m$ holds for any $k\ge0$ and $i = 1,\cdots,n$. If Assumption \ref{assum:g} holds, then for any integers $k\ge0$ and $h\ge1$, we have
	\begin{align*}
	\lambda_{\min}\left(\sum_{j=k+1}^{k+h}\left({\rm diag}\{A_{j,1},\cdots,A_{j,n} \}+\nu\mathscr{L}\right)\right)
	\ge \frac{\ell_{m+1}\nu}{2nc+\ell_{m+1}\nu n} \lambda_{\min}\left(\sum_{i =1}^n\sum_{j = k+1}^{k+h}A_{j,i} \right),
	\end{align*}
	where $\ell_{m+1}$ denotes the $(m+1)$-th smallest eigenvalue of $\mathscr{L}$ in (\ref{z}) (i.e., $\lambda_2(\mathcal{L})$ in Remark \ref{rmk:2}).
\end{lemma}

\begin{assumption}(Conditional-expectation-based Cooperative Excitation Condition)\label{excit:cc}
	For the adapted regression sequence $\{x_{k,i}, \mathscr{F}_k\}$, where $\mathscr{F}_k \triangleq \sigma\{x_{j,i}, q_{j+1,i}, d_{j-1,i}, \omega_j,$ $i = 1,\cdots,n, j \le k \}$ is a non-decreasing $\sigma$-algebra, there exist an integer $h \in\mathbb{N}^+$ and an excitation constant $\beta \in (0, 1)$ such that, for any $t \ge 1$,
	\begin{equation*}
	\mathbb{E}\left[\frac{1}{n(1+h)}\sum_{i=1}^n\sum_{k=t+1}^{t+h}\frac{x_{k,i}x_{k,i}^\top}{1+\|x_{k,i}\|^2}\Bigg|\mathscr{F}_{t}\right]\ge\beta I_m>0, ~{\rm{a.s.}}
	\end{equation*}
\end{assumption}

\begin{remark}
Assumption \ref{excit:cc} is the conditional expectation type cooperative excitation condition for stochastic regressors.  Compared with the deterministic excitation conditions such as those in \cite{Wang2022Unified,javed2022}, it allows greater temporal and statistical dependence in the regression process.
This type of conditional-expectation-based excitation condition was first  introduced in \cite{Guo1990Estimating} for the stability analysis of Kalman filtering in a single-sensor setup, which takes the form:
	\begin{equation}
	\mathbb{E}\left[\frac{1}{(1+h)}\sum_{k=t+1}^{t+h}\frac{x_{k,i}x_{k,i}^\top}{1+\|x_{k,i}\|^2}\Bigg|\mathscr{F}_{t}\right]\ge\alpha I_m>0, \quad{\rm{a.s..}}\label{single}
\end{equation}
In contrast to the classical single-sensor criterion, Assumption \ref{excit:cc} incorporates both temporal evolution information and spatial collaborative information among all sensors.  Consequently,  Assumption \ref{excit:cc} is less restrictive than the condition (\ref{single}), and reduces to the classical counterpart (\ref{single}) when $n = 1$. Unlike \cite{Carpentiero-2024}, no independence or stationarity conditions are imposed on the regressors $\{x_{k,i}\}$. Thus, it allows stochastic regression signals arising from feedback loops, and includes a general class of stochastic processes such as asymptotically independent processes and those generated by linear dynamical models, see e.g., \cite{Guo1990Estimating, Wang2026Distributed}.
\end{remark}

\begin{lemma}\label{lem:plus}
Under Assumptions \ref{assum:g} and \ref{excit:cc}, the matrix $G_k$ defined in (\ref{z}) satisfies the following property:
\begin{equation}\label{excit:2}
\lambda_{\min}\left\{\mathbb{E}\left[\sum_{j=k+1}^{k+h}G_j\Bigg|\mathscr{F}_k\right] \right\}\ge\frac{\ell_{m+1}\nu(h+1)\beta}{2+\ell_{m+1}\nu }  ,\quad{\rm{a.s.}}
\end{equation}
where $\beta$, $h$, and $\mathscr{F}_{k}$ are given in Assumption \ref{excit:cc}, $\ell_{m+1}$ is  defined in Lemma \ref{lem:sigma}.
\end{lemma}

\begin{proof}
	Let $A_{j,i} \triangleq \mathbb{E}\left[\frac{x_{j,i}x_{j,i}^\top}{1+\|x_{j,i}\|^2}\Big|\mathscr{F}_k\right]$. Then, we have $0\le A_{j,i}\le I_m$. By Lemma \ref{lem:sigma}, 
    we can further obtain
	\begin{equation*}
	\lambda_{\min}\left\{\mathbb{E}\left[\sum_{j=k+1}^{k+h}G_j\Bigg|\mathscr{F}_k\right] \right\} \ge \frac{\ell_{m+1}\nu}{2n+\ell_{m+1}\nu n} \lambda_{\min}\left(\mathbb{E}\left[\sum_{i=1}^n\sum_{j=k+1}^{k+h}\frac{x_{j,i}x_{j,i}^\top}{1+\|x_{j,i}\|^2}\Bigg|\mathscr{F}_k\right] \right).
	\end{equation*}
	The conclusion then holds by combining this result with Assumption \ref{excit:cc}.
\end{proof}


Besides Assumption \ref{assum:1} on the compression operator, the stability analysis also requires conditions on the observation noise and parameter variations. Specifically, they are assumed to satisfy martingale-difference and moment-boundedness conditions.

\begin{assumption}\label{assum:5}
	For any $k\ge1$ and $1\le i\le n$, the observation noise $d_{k,i}$ and parameter variation $\omega_{k+1}$ satisfy
	\begin{align*}
	&\mathbb{E}[d_{k,i}|\mathscr{F}_k] = 0,\quad\mathbb{E}[ \omega_{k+1}|\mathscr{F}_k] = 0,\quad \mathbb{E}[d_{k,i} \omega_{k+1}|\mathscr{F}_k] = 0,\\
	&\sup_{k\ge1}\mathbb{E}\| D_k\|^2\le c_{d}<\infty,\quad \sup_{k\ge1}\mathbb{E}\|\Omega_{k+1}\|^2\le c_{\omega}<\infty,
	\end{align*}	
	where $\mathscr{F}_{k}$ is given in Assumption \ref{excit:cc}.
\end{assumption}

\begin{remark}\label{rmk:3}
    The compression error $e_{k+1,i}= Q(\delta_{k+1,i})-\delta_{k+1,i}$ is conditionally unbiased and has zero conditional cross-moments with the observation noise and parameter variation, which is useful for the subsequent mean-square analysis.
    To be specific, from the recursion (\ref{recur:1}) of $\delta_{k+1,i}$, we know that $\delta_{k+1,i}\in\sigma\{\mathscr{F}_k,d_{k,i},\omega_{k+1}\}\triangleq\bar{\mathscr{F}}_k$. 
    Given $\bar{\mathscr{F}}_k$ (any given $\hat{\xi}_{0:k+1,i}$ and $q_{0:k+1,i}$), the randomness of compression operator $Q(\delta_{k+1,i})$ is generated solely by the compression operator for the fixed input  $\delta_{k+1,i}$. Hence, by Assumption \ref{assum:1}, Remark \ref{re1} and equation (\ref{co:3}), we have
	\begin{align*}
		\mathbb{E}[e_{k+1,i}|\mathscr{F}_k]&=\mathbb{E}\left[\mathbb{E}[e_{k+1,i}|\bar{\mathscr{F}}_k]\big|\mathscr{F}_k\right] 
		=\mathbb{E}\left[\mathbb{E}[e_{k+1,i}|\delta_{k+1,i}]\big|\mathscr{F}_k\right]=0.
	\end{align*}
	   Using the same argument, the cross-terms between the compression error $e_{k+1,i}$ and the observation noise $d_{k,i}$ or the parameter variation $\omega_{k+1}$ satisfy
	\begin{align*}	\mathbb{E}\left[d_{k,i}e_{k+1,i}|\mathscr{F}_k \right] 
    =0,\quad
	\mathbb{E}\left[\omega_{k+1}^\top e_{k+1,i}|\mathscr{F}_k \right] 
    =0.
	\end{align*}
\end{remark}
Based on the above assumptions and lemmas, we first present the mean-square stability of the auxiliary compression variable error $\widetilde{q}_k$.

\begin{theorem}\label{thm:consensus}
	Consider the error equation (\ref{equ:2}) of Algorithm \ref{algo:1} under Assumptions \ref{assum:1}-\ref{assum:5}. 
    Suppose that 
	the compression weight $\zeta$ and the compression parameter $\rho$ satisfy  $\zeta(1+\rho)<2$. 
    If $\mu \in (0, 1)$ and $\nu \in (0, 1)$ satisfy $\mu (1 + 2\nu ) \leq 1$, the auxiliary compression error $\widetilde{q}_k$ is mean-square bounded and satisfies
	\begin{align*}
	\limsup_{k \to \infty}\mathbb{E}\|\widetilde{ q}_k\|^2&\le \frac{c_2c_d+c_3c_\omega}{c_4},
	\end{align*}
	where $c_d$ and $c_\omega$ are defined in Assumption \ref{assum:5},
	\begin{equation*}
	\begin{aligned}
	&c_0 = \zeta\left[2-\zeta(1+\rho)\right]\mu,~~~c_1 = 4\zeta^2(1+\rho)h(h+1),\\
	&c_2 = \zeta^2(1+\rho)\mu^2(h+1)+\frac{c_0c_1\mu(h+1)}{2+c_1},\\
	&c_3 = \gamma^2\left[h+1+\frac{4c_0h(h+1)^2}{(2+c_1)\mu} \right],\\
	&c_4 = \frac{\ell_{m+1}\nu(h+1)\beta c_0}{(2+\ell_{m+1}\nu)(2+c_1)},
	\end{aligned}
	\end{equation*}
	and $\beta,h, \ell_{m+1}$ are the same as those  in Lemma \ref{lem:plus}.
\end{theorem}
The proof of Theorem \ref{thm:consensus} is provided in Subsection \ref{proof:theorem1}.
Using Theorem \ref{thm:consensus} together with the recursion (\ref{q-xi}), we next derive an upper bound on the tracking error $\widetilde{\Xi}_{k+1}$ and thus establish the stability of Algorithm \ref{algo:1}.

\begin{theorem}\label{thm:11}
Under the conditions of Theorem \ref{thm:consensus}, the tracking error  $\widetilde{\Xi}_k$ of Algorithm \ref{algo:1} is mean-square bounded and satisfies
	$$
	\limsup_{k \to \infty} \mathbb{E}\|\widetilde{\Xi}_k\|^2 \leq \left(\frac{c_2}{c_4} + \mu^2\right)c_d + \left(\frac{c_3}{c_4} + \gamma^2\right)c_\omega.
	$$
	where   $c_2$ $c_3$ and $c_4$ are defined in Theorem \ref{thm:consensus}, and  $c_d$, $c_\omega$ are defined in Assumption \ref{assum:5}.
\end{theorem}

The proof of Theorem \ref{thm:11} is provided in Subsection \ref{proof:theorem2}.
\begin{remark}\label{rmk:a} 
    The error bounds in Theorems \ref{thm:consensus} and \ref{thm:11} characterize the effects of observation noise and parameter variation on the steady-state tracking performance. 
    In particular, in the noise-free and time-invariant case, i.e., \(c_d=0\) and \(\gamma=0\), the derived error bounds vanish. This indicates that the proposed compression mechanism does not introduce a nonzero asymptotic estimation error in this special case, while still reducing communication cost.
\end{remark}

\begin{remark}\label{remark6}
The derived bounds reveal how the algorithm, excitation, network, and compression parameters affect the tracking performance. 
From Theorems \ref{thm:consensus} and \ref{thm:11}, the coefficients associated with observation noise and parameter variation are proportional to
\[
\frac{c_2}{c_4}
=\frac{\mu(2+\ell_{m+1}\nu)}{\ell_{m+1}\nu\beta}
\frac{2\zeta(1+\rho)+8\zeta^2(1+\rho)h(h+1)}
{2-\zeta(1+\rho)},
\]
and
\[
\frac{c_3}{c_4}
=\frac{\gamma^2(2+\ell_{m+1}\nu)}
{\ell_{m+1}\nu\beta\mu}
\frac{2+8\zeta h(h+1)}
{\zeta[2-\zeta(1+\rho)]},
\]
respectively. 
Hence, increasing \(\mu\) increases the noise-induced error but reduces the parameter-variation-induced error. This reveals a trade-off between noise sensitivity and tracking ability.
For fixed feasible \(\mu\), a larger consensus weight \(\nu\) tends to improve the bound. 
Stronger excitation, reflected by larger \(\beta\) and smaller \(h\), and better network connectivity, characterized by larger \(\ell_{m+1}\), also improve the tracking performance.
The compression parameter \(\rho\) characterizes the relative compression error.
A larger $\rho$ means a higher degree of data compression, but leads to a larger error bound.
The effect of the compression weight \(\zeta\) is nonmonotonic for the parameter-variation term, whose coefficient is minimized at
\[
\zeta^\ast=
\frac{
\sqrt{(1+\rho)^2+8h(h+1)(1+\rho)}-(1+\rho)}
{4h(h+1)(1+\rho)}.
\]

\end{remark}

\section{Proofs of the main results}\label{proofs_of}

\subsection{ Proof of Theorem \ref{thm:consensus}}\label{proof:theorem1}
\begin{proof}
By the definitions of $\widetilde{q}_{k,i}$ in (\ref{q}) and  $\mathscr{F}_{k}$ in Assumption \ref{excit:cc}, we know that $\widetilde{q}_{k}\in\mathscr{F}_{k}$. Thus,
based on the recursive relation of $\widetilde{q}_{k+1}$ in Equation (\ref{equ:2}), combined with Assumption \ref{assum:5} and Remark \ref{rmk:3}, we obtain
\begin{align}
\mathbb{E}\left[\widetilde{ q}_{k+1}^\top\widetilde{ q}_{k+1}|\mathscr{F}_k \right] ={}& \mathbb{E}\left[\widetilde{ q}_{k}^\top\left(I-\zeta\mu G_k\right)^\top\left(I-\zeta\mu G_k\right)\widetilde{ q}_{k}\Big|\mathscr{F}_k \right]+\zeta^2\mathbb{E}\left[ e_{k+1}^\top  e_{k+1}|\mathscr{F}_k \right]\notag\\
&+\zeta^2\mu^2\mathbb{E}\left[ \|L_k D_k\|^2|\mathscr{F}_k \right]+\gamma^2\mathbb{E}\left[\| \Omega_{k+1}\|^2|\mathscr{F}_k \right].\label{nnn20}
\end{align}
By Assumption \ref{assum:1} and Remark \ref{re1} (see (\ref{co:4})), it follows from the analysis method outlined in Remark \ref{rmk:3} that
\begin{align}
\mathbb{E}\left[ e_{k+1}^\top  e_{k+1}|\mathscr{F}_k \right] ={}& \mathbb{E}\left[\mathbb{E}[ e_{k+1}^\top  e_{k+1}|\bar{\mathscr{F}}_k]\big|\mathscr{F}_k \right]\le \rho \mathbb{E}\left[ \delta_{k+1}^\top  \delta_{k+1}|\mathscr{F}_k \right]\notag \\
={}&\rho\mu^2\mathbb{E}\left[\widetilde{ q}_{k}^\top G_k^\top G_k \widetilde{ q}_{k}|\mathscr{F}_k \right]+\rho\mu^2\mathbb{E}\left[ \|L_k D_k\|^2|\mathscr{F}_k \right],\label{eqn:6}
\end{align}
where the last equality follows from the recursive relation of $\delta_{k+1}$ in (\ref{e2}).
Furthermore, by the symmetry of $G_k$ and Lemma \ref{lem:e}, we have from (\ref{nnn20})
\begin{align}
\mathbb{E}\left[\widetilde{ q}_{k+1}^\top\widetilde{ q}_{k+1}|\mathscr{F}_k \right]
\le{}& \widetilde{ q}_{k}^\top\widetilde{ q}_{k}-\widetilde{ q}_{k}^\top\left[2\zeta\mu G_k-\zeta^2(1+\rho)\mu^2 G_k^2 \right]\widetilde{ q}_k+\zeta^2(1+\rho)\mu^2\mathbb{E}\left[ \| L_k D_k\|^2|\mathscr{F}_k \right]+\gamma^2\mathbb{E}\left[\| \Omega_{k+1}\|^2|\mathscr{F}_k \right]\nonumber\\
\le{}& \widetilde{ q}_{k}^\top\widetilde{ q}_{k}-\zeta\mu\left[2-\zeta(1+\rho)\right]\widetilde{ q}_{k}^\top G_k \widetilde{ q}_{k}+\zeta^2(1+\rho)\mu^2\mathbb{E}\left[ \|L_k D_k\|^2|\mathscr{F}_k \right]+\gamma^2\mathbb{E}\left[\| \Omega_{k+1}\|^2|\mathscr{F}_k \right].\label{eq23}
\end{align}
By applying the conditional expectation $\mathbb{E}[\cdot|\mathscr{F}_{k-1}]$ to both sides of the above expression, we obtain that
\begin{align*}
\mathbb{E}\left[\widetilde{ q}_{k+1}^\top\widetilde{ q}_{k+1}|\mathscr{F}_{k-1} \right]
\le{}& \mathbb{E}[\widetilde{ q}_{k}^\top\widetilde{ q}_{k}|\mathscr{F}_{k-1}]-\zeta\mu\left[2-\zeta(1+\rho)\right]\mathbb{E}[\widetilde{ q}_{k}^\top G_k \widetilde{ q}_{k}|\mathscr{F}_{k-1}]\\
&+\zeta^2(1+\rho)\mu^2\mathbb{E}\left[ \|L_k D_k\|^2|\mathscr{F}_{k-1} \right]+\gamma^2\mathbb{E}\left[\| \Omega_{k+1}\|^2|\mathscr{F}_{k-1} \right]\\
\le{}& \widetilde{ q}_{k-1}^\top\widetilde{ q}_{k-1}-\zeta\mu\left[2-\zeta(1+\rho)\right]\left(\widetilde{ q}_{k-1}^\top G_{k-1} \widetilde{ q}_{k-1}+\mathbb{E}[\widetilde{ q}_{k}^\top G_k \widetilde{ q}_{k}|\mathscr{F}_{k-1}]\right)\\
&+\zeta^2(1+\rho)\mu^2\mathbb{E}\left[ \|L_{k-1 }D_{k-1}\|^2+\|L_k D_k\|^2|\mathscr{F}_{k-1} \right]+\gamma^2\mathbb{E}\left[\| \Omega_{k}\|^2+\| \Omega_{k+1}\|^2|\mathscr{F}_{k-1} \right].
\end{align*}
Through $h$ more recursive steps, it can be derived that
\begin{align}
\mathbb{E}\left[\widetilde{ q}_{k+1}^\top\widetilde{ q}_{k+1}|\mathscr{F}_{k-h} \right]
\le{}&\widetilde{ q}_{k-h}^\top\widetilde{ q}_{k-h}-c_0\mathbb{E}\left[\sum_{j=1}^{h+1}\widetilde{ q}_{k-j+1}^\top G_{k-j+1} \widetilde{ q}_{k-j+1}\Bigg|\mathscr{F}_{k-h}\right]\notag\\
&+\zeta^2(1+\rho)\mu^2\mathbb{E}\left[\sum_{j=1}^{h+1}\left\| L_{k-j+1} D_{k-j+1} \right\|^2\Bigg|\mathscr{F}_{k-h}\right]+\gamma^2\mathbb{E}\left[\sum_{j=1}^{h+1}\| \Omega_{k-j+2}\|^2\Bigg|\mathscr{F}_{k-h} \right],\label{recur:4}
\end{align}
where $c_0 = \zeta\mu\left[2-\zeta(1+\rho)\right]$.

We now focus our analysis on the second term on the right-hand side of  (\ref{recur:4}). First, we examine the explicit expression of $\mathbb{E}[\left\|\widetilde{ q}_{k-j+1}-\widetilde{ q}_{k-h}\right\|^2|\mathscr{F}_{k-h} ]$ $(1\le j\le h+1)$, noting that by (\ref{equ:2})
\begin{align*}
\widetilde{ q}_{k-j+1}
={}&\widetilde{ q}_{k-j}-\zeta\mu G_{k-j}\widetilde{ q}_{k-j}+\zeta e_{k-j+1}+\zeta\mu L_{k-j}D_{k-j}-\gamma\Omega_{k-j+1}\\
={}& \widetilde{ q}_{k-h}-\sum_{i = j+1}^{h+1}\zeta\mu G_{k-i+1}\widetilde{ q}_{k-i+1}+\sum_{i = j+1}^{h+1}\left(\zeta e_{k-i+2}+\zeta\mu L_{k-i+1}D_{k-i+1}-\gamma\Omega_{k-i+2}\right).
\end{align*}
Then by the $C_r$-inequality, it yields
\begin{align}
\mathbb{E}\left[\left\|\widetilde{ q}_{k-j+1}-\widetilde{ q}_{k-h}\right\|^2\Big|\mathscr{F}_{k-h} \right]
\le{}&4\zeta^2\mu^2\mathbb{E}\left[\left\|\sum_{i=j+1}^{h+1}G_{k-i+1}\widetilde{ q}_{k-i+1}\right\|^2\Bigg|\mathscr{F}_{k-h} \right]+4\zeta^2\mathbb{E}\left[\left\|\sum_{i=j+1}^{h+1}e_{k-i+2}\right\|^2\Bigg|\mathscr{F}_{k-h} \right]\notag\\
&+4\zeta^2\mu^2\mathbb{E}\left[\left\|\sum_{i=j+1}^{h+1}L_{k-i+1} D_{k-i+1}\right\|^2\Bigg|\mathscr{F}_{k-h} \right]+4\gamma^2\mathbb{E}\left[\left\|\sum_{i=j+1}^{h+1} \Omega_{k-i+2}\right\|^2\Bigg|\mathscr{F}_{k-h} \right].\label{eqn:11}
\end{align}
Furthermore, it follows from equation (\ref{eqn:6}) that
\begin{align}
\mathbb{E}\left[\left\|\sum_{i=j+1}^{h+1}e_{k-i+2}\right\|^2\Bigg|\mathscr{F}_{k-h} \right]
\le{}&(h-j+1)\mathbb{E}\left[\sum_{i=j+1}^{h+1}\left\|e_{k-i+2}\right\|^2\Bigg|\mathscr{F}_{k-h} \right]\notag\\
\le{}&\rho\mu^2(h-j+1)\mathbb{E}\left[\sum_{i=j+1}^{h+1}\left(\left\|G_{k-i+1}\widetilde{ q}_{k-i+1}\right\|^2+\left\| L_{k-i+1} D_{k-i+1}\right\|^2\right)\Bigg|\mathscr{F}_{k-h} \right].\label{eqn:22}
\end{align}
Therefore, combining equations (\ref{eqn:11}) and (\ref{eqn:22}), we have
\begin{align}
\mathbb{E}\left[\left\|\widetilde{ q}_{k-j+1}-\widetilde{ q}_{k-h}\right\|^2|\mathscr{F}_{k-h} \right]
\le{}&4\zeta^2(1+\rho)\mu^2(h-j+1)\mathbb{E}\left[\sum_{i=j+1}^{h+1}\left(\left\|G_{k-i+1}\widetilde{ q}_{k-i+1}\right\|^2+\left\| L_{k-i+1} D_{k-i+1}\right\|^2\right)\Bigg|\mathscr{F}_{k-h} \right]\nonumber\\
&+4\gamma^2(h-j+1)\mathbb{E}\left[\sum_{i=j+1}^{h+1}\left\| \Omega_{k-i+2}\right\|^2\Bigg|\mathscr{F}_{k-h} \right].\label{e21}
\end{align}
According to Lemma \ref{lem:e}, we can get $\left\|G_{k-i+1}\widetilde{ q}_{k-i+1}\right\|^2\le\frac{1}{\mu}\widetilde{ q}_{k-i+1}^\top G_{k-i+1}\widetilde{ q}_{k-i+1}$ and $\|G_{k-j+1}\|\le\frac{1}{\mu}$.

So it can be further derived from (\ref{e21}) that
\begin{align}
&\mathbb{E}\left[\sum_{j=1}^{h+1}\left\|G_{k-j+1}^{\frac{1}{2}}\left(\widetilde{ q}_{k-j+1}-\widetilde{ q}_{k-h}\right)\right\|^2\Bigg|\mathscr{F}_{k-h} \right]
\le\frac{1}{\mu}\mathbb{E}\left[\sum_{j=1}^{h+1}\left\|\widetilde{ q}_{k-j+1}-\widetilde{ q}_{k-h}\right\|^2\Bigg|\mathscr{F}_{k-h} \right]\notag\\
\le{}&4\zeta^2(1+\rho)\mu\sum_{j=1}^{h+1}(h-j+1)\mathbb{E}\left[\sum_{i=j+1}^{h+1}\left(\left\|G_{k-i+1}\widetilde{ q}_{k-i+1}\right\|^2+\left\| L_{k-i+1} D_{k-i+1}\right\|^2\right)\Bigg|\mathscr{F}_{k-h} \right]\nonumber\\
&+\frac{4}{\mu}\gamma^2\sum_{j=1}^{h+1}(h-j+1)\mathbb{E}\left[\sum_{i=1}^{h+1}\left\| \Omega_{k-i+2}\right\|^2\Bigg|\mathscr{F}_{k-h} \right]\nonumber\\
\le{}&2\zeta^2(1+\rho) h(h+1)\mathbb{E}\left[\sum_{i=1}^{h+1}\left\|G_{k-i+1}^\frac{1}{2}\widetilde{ q}_{k-i+1}\right\|^2\Bigg|\mathscr{F}_{k-h} \right]+2\zeta^2(1+\rho)\mu h(h+1)\mathbb{E}\left[\sum_{i=1}^{h+1}\left\| L_{k-i+1} D_{k-i+1}\right\|^2\Bigg|\mathscr{F}_{k-h} \right]\nonumber\\
&+\frac{2}{\mu}\gamma^2h(h+1)\mathbb{E}\left[\sum_{i=1}^{h+1}\left\| \Omega_{k-i+2}\right\|^2\Bigg|\mathscr{F}_{k-h} \right].\label{e3}
\end{align}
Then by (\ref{e3}), we have
\begin{align}
&\mathbb{E}\left[\sum_{j=1}^{h+1}\left\|G_{k-j+1}^{\frac{1}{2}}\widetilde{ q}_{k-h}\right\|^2\Bigg|\mathscr{F}_{k-h} \right]
\le2\mathbb{E}\left[\sum_{j=1}^{h+1}\left(\left\|G_{k-j+1}^{\frac{1}{2}}\left(\widetilde{ q}_{k-j+1}-\widetilde{ q}_{k-h}\right)\right\|^2+\left\|G_{k-j+1}^{\frac{1}{2}}\widetilde{ q}_{k-j+1}\right\|^2\right)\Bigg|\mathscr{F}_{k-h} \right]\notag\\
\le{}& \left(2+c_1\right)\mathbb{E}\left[\sum_{j=1}^{h+1}\left\|G_{k-j+1}^{\frac{1}{2}}\widetilde{ q}_{k-j+1}\right\|^2\Bigg|\mathscr{F}_{k-h} \right]+c_1\mu\mathbb{E}\left[\sum_{j=1}^{h+1}\left\| L_{k-j+1} D_{k-j+1}\right\|^2\Bigg|\mathscr{F}_{k-h} \right]\notag\\
&+\frac{4}{\mu}\gamma^2h(h+1)\mathbb{E}\left[\sum_{j=1}^{h+1}\left\| \Omega_{k-j+2}\right\|^2\Bigg|\mathscr{F}_{k-h} \right],\label{eqn:44}
\end{align}
where $c_1 = 4\zeta^2(1+\rho)h(h+1)$.

Thus, equation (\ref{eqn:44}) can be equivalently represented as
\begin{align}
&\mathbb{E}\left[\sum_{j=1}^{h+1}\widetilde{ q}_{k-j+1}^\top G_{k-j+1} \widetilde{ q}_{k-j+1}\Bigg|\mathscr{F}_{k-h}\right]\\
\ge&\frac{1}{2+c_1}\mathbb{E}\left[\sum_{j=1}^{h+1}\left\|G_{k-j+1}^{\frac{1}{2}}\widetilde{ q}_{k-h}\right\|^2\Bigg|\mathscr{F}_{k-h} \right]-\frac{c_1\mu}{2+c_1}\mathbb{E}\left[\sum_{j=1}^{h+1}\left\|L_{k-j+1} D_{k-j+1}\right\|^2\Bigg|\mathscr{F}_{k-h} \right]\notag\\
&-\frac{4\gamma^2h(h+1)}{(2+c_1)\mu}\mathbb{E}\left[\sum_{j=1}^{h+1}\left\| \Omega_{k-j+2}\right\|^2\Bigg|\mathscr{F}_{k-h} \right].\label{eqn:7}
\end{align}
Substituting (\ref{eqn:7}) into (\ref{recur:4}), we have
\begin{align*}
\mathbb{E}\left[\widetilde{ q}_{k+1}^\top\widetilde{ q}_{k+1}|\mathscr{F}_{k-h} \right]
\le{}&\widetilde{ q}_{k-h}^\top\widetilde{ q}_{k-h}-\frac{c_0}{2+c_1}\widetilde{ q}_{k-h}^\top\mathbb{E}\left[\sum_{j=1}^{h} G_{k-j+1} \Bigg|\mathscr{F}_{k-h}\right]\widetilde{ q}_{k-h}+R_k\\
\le{}&\left(1-\frac{c_0}{2+c_1}\frac{\ell_{m+1}\nu(h+1)\beta}{2+\ell_{m+1}\nu}\right)\widetilde{ q}_{k-h}^\top\widetilde{ q}_{k-h}+R_k,
\end{align*}
where
\begin{equation*}
R_k\triangleq\left[\zeta^2(1+\rho)\mu^2\!+\!\frac{c_0c_1\mu}{2+c_1}\right]\mathbb{E}\left[\sum_{j=1}^{h+1}\left\| L_{k-j+1} D_{k-j+1} \right\|^2\Bigg|\mathscr{F}_{k-h}\right]\!+\!\gamma^2\left[1\!+\!\frac{4c_0h(h+1)}{(2+c_1)\mu} \right]\mathbb{E}\left[\sum_{j=1}^{h+1}\| \Omega_{k-j+2}\|^2\Bigg|\mathscr{F}_{k-h} \right].
\end{equation*}
The final inequality follows from Lemma \ref{lem:plus}. By taking expectations on both sides, we can obtain that
\begin{align}\label{qici}
\mathbb{E}\left[\widetilde{ q}_{k+1}^\top\widetilde{ q}_{k+1} \right]
\le\left(1-\frac{\ell_{m+1}\nu(h+1)\beta c_0}{(2+\ell_{m+1}\nu)(2+c_1)}\right)\mathbb{E}[\widetilde{ q}_{k-h}^\top\widetilde{ q}_{k-h}]+c_2c_{d}+c_3c_\omega,
\end{align}
where
\begin{align*}
&c_0 = \zeta\left[2-\zeta(1+\rho)\right]\mu,\quad c_1 = 4\zeta^2(1+\rho)h(h+1),\\
&c_2 = \zeta^2(1+\rho)\mu^2(h+1)+\frac{c_0c_1\mu(h+1)}{2+c_1},\quad c_3 = \gamma^2\left[h+1+\frac{4c_0h(h+1)^2}{(2+c_1)\mu} \right],
\end{align*}
and $\beta,h,\mathscr{F}_{k-h},\ell_{m+1}$ are the same as those in Lemma \ref{lem:plus}.

We now proceed to analyze the homogeneous term in Equation (\ref{qici}) in detail. If the compression weight $\zeta$ of Algorithm \ref{algo:1} and the compression parameter $\rho$ of the random compression operator satisfy $\zeta(1+\rho) < 2$, then
\begin{align*}
&\frac{\ell_{m+1}\nu(h+1)\beta \zeta\mu\left[2-\zeta(1+\rho)\right]}{(2+\ell_{m+1}\nu)(2+4(1+\rho)\zeta^2h(h+1))}
\le \frac{\ell_{m+1}\nu(h+1)\beta \zeta\mu\left[2-\zeta(1+\rho)\right]}{(2+\ell_{m+1}\nu)4\sqrt{2(1+\rho)h(h+1)}\zeta }\\
=&\frac{\ell_{m+1}\nu\sqrt{h+1}\beta \mu\left[2-\zeta(1+\rho)\right]}{(2+\ell_{m+1}\nu)4\sqrt{2(1+\rho)h} }
<\frac{\sqrt{h+1}\beta \mu\left[2-\zeta(1+\rho)\right]}{8\sqrt{2(1+\rho)h} }
<\frac{\sqrt{h+1}\beta \mu}{4\sqrt{2(1+\rho)h} }
\le\frac{\beta \mu}{4\sqrt{1+\rho} }<\frac{1}{4},
\end{align*}
where the first inequality follows from the mean value inequality, while the second one holds due to the parameter ranges $\ell_{m+1} \in (0, 2)$ and $\nu \in (0, 1)$.

Therefore, if $\zeta(1+\rho)<2$ holds, we have
\begin{equation}\label{c4}
1-\frac{\ell_{m+1}\nu(h+1)\beta c_0}{(2+\ell_{m+1}\nu)(2+c_1)}\triangleq1-c_4\in(3/4,1).
\end{equation}
Since the recurrence relation (\ref{qici}) operates with a time interval of $h+1$, without loss of generality, consider any time $k = (h+1)l + k_0$, where $l$ is a positive integer and $k_0 \in \{0, 1, \ldots, h\}$. Then, we have
\begin{align}
\mathbb{E}\left[\widetilde{ q}_{(h+1)l+k_0}^\top\widetilde{ q}_{(h+1)l+k_0} \right]
\le{}&\left(1-c_4\right)\mathbb{E}\left[\widetilde{ q}_{(h+1){(l-1)}+k_0}^\top\widetilde{ q}_{(h+1)(l-1)+k_0}\right]+c_2c_{d}+c_3c_\omega\notag\\
\le{}&\left(1-c_4\right)^2\mathbb{E}\left[\widetilde{ q}_{(h+1){(l-2)}+k_0}^\top\widetilde{ q}_{(h+1)(l-2)+k_0}\right]+(c_2c_{d}+c_3c_\omega)(1+1-c_4) \notag\\  \le&\cdots\notag\\
\le{}&\left(1-c_4\right)^l\mathbb{E}\left[\widetilde{ q}_{k_0}^\top\widetilde{ q}_{k_0}\right]+(c_2c_{d}+c_3c_\omega)\left(1+1-c_4+(1-c_4)^{l-1}\right)\notag\\
={}&\left(1-c_4\right)^l\mathbb{E}\left[\widetilde{ q}_{k_0}^\top\widetilde{ q}_{k_0}\right]+(c_2c_{d}+c_3c_\omega)\frac{1-(1-c_4)^l}{c_4}.\label{speed}
\end{align}
Taking the upper limit with respect to $l$ in the above expression yields the conclusion of the theorem.
\end{proof}

\subsection{Proof of Theorem \ref{thm:11} } \label{proof:theorem2}
\begin{proof}
	Combining  (\ref{q-xi}) and  Theorem \ref{thm:consensus}, it can be obtained from Assumption \ref{assum:5} that
	\begin{align*}
	\limsup_{k \to \infty}\mathbb{E}\|\widetilde{\Xi}_k\|^2
	&= \limsup_{k \to \infty}\mathbb{E}\left\|(I-\mu G_k)\widetilde{q}_{k}+\mu L_kD_k-\gamma\Omega_{k+1}\right\|^2\\
	&\le\limsup_{k \to \infty}\mathbb{E}\left[\left\|(I-\mu G_k)\widetilde{q}_{k}\right\|^2+\mu^2\left\| L_kD_k\right\|^2+\gamma^2\left\|\Omega_{k+1}\right\|^2\right]\\
	&\le\limsup_{k \to \infty}\mathbb{E}\left[\left\|\widetilde{q}_{k}\right\|^2+\mu^2\left\|D_k\right\|^2+\gamma^2\left\|\Omega_{k+1}\right\|^2\right]= \left(\frac{c_2}{c_4}+\mu^2\right)c_d+\left(\frac{c_3}{c_4}+\gamma^2\right)c_\omega.
	\end{align*}
	This completes the proof.
\end{proof}

\section{A simulation example}\label{simulaiton}
This section provides numerical simulations to illustrate the theoretical results of the proposed CC-DLMS algorithm over a fixed undirected network, with a focus on the effects of communication compression and parameter variation.

\begin{figure}[!htb]
	\centering
	\includegraphics[width=3.5in]{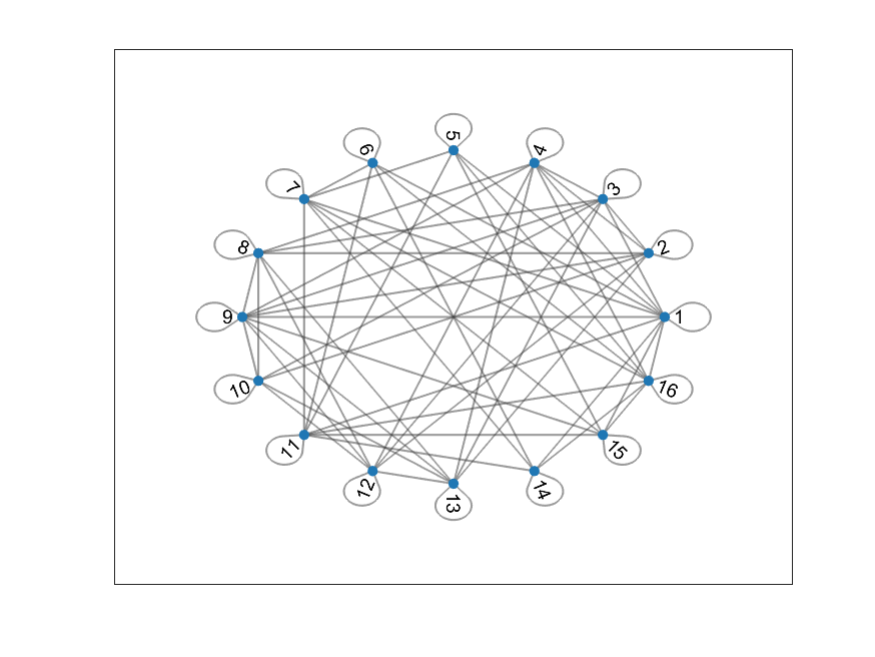}
    \vspace{-10pt}
	\caption{Network topology of 16 sensors.}\label{Graph}
\end{figure}

Consider a sensor network consisting of $n=16$ sensors, where each sensor collects observations according to the stochastic regression model (\ref{e:xt}). The dimension of the unknown time-varying parameter vector is set to $m=80$.
The communication topology is generated as an undirected graph, where the weighted adjacency matrix is constructed via the Metropolis rule to guarantee column stochasticity. The resulting network topology is depicted in Figure \ref{Graph}, which satisfies the connectivity condition in Assumption \ref{assum:g}. 

For each sensor \(i \in \{1,\ldots,n\}\), the regressor
\(x_{k,i} \in \mathbb{R}^{m}\) is generated by
$
x_{k,i}
=
0.9^{k}\mathbf{1}_{m}
+
\cos\left(\frac{i\pi}{n}\right)s_{k,i},
$
where \(\mathbf{1}_{m}\) denotes the \(m\)-dimensional all-ones
vector, and \(\{s_{k,i}\}_{k\geq 0}\) is an \(m\)-dimensional
first-order autoregressive process:
$
s_{k,i}
=
0.9s_{k-1,i}+\nu_{k,i},
\quad
s_{-1,i}=\mathbf{0}.
$
Here, \(\nu_{k,i}=0.4u_{k,i}\), where the components of \(u_{k,i}\in\mathbb{R}^{m}\) are independent random variables, each taking the values \(+1\) and \(-1\) with equal probability.
It can be verified that the regression vectors $x_{k,i}$ satisfy the cooperative excitation condition in Assumption \ref{excit:cc}.
The observation noise $d_{k,i}$  is zero-mean white Gaussian with variance $\sigma_d^2 = 0.01$, independent across sensors and time instants.
The unknown parameter evolves as (\ref{div})
where $\omega_k \sim \mathcal{N}(0, I_m)$ and $\gamma$ controls the parameter variation rate. Two scenarios are investigated: time-invariant parameters ($\gamma = 0$) and slowly time-varying parameters ($\gamma = 0.003$).

To guarantee a fair comparison, all algorithms adopt identical initial conditions. 
Specifically, the local estimates of all sensors are initialized as zero vectors, i.e., $\hat{\xi}_{0,i}=\mathbf{0}$ for all $i=1,2,\cdots,n$, and the auxiliary compression variables in CC-DLMS are also initialized to zero. 
The adaptive step size and consensus weight are set to $\mu = 0.7$ and $\nu = 0.2$, respectively, which satisfy the condition $\mu(1+2\nu) \leq 1$. 
The randomized quantization operator $Q(\cdot)$ introduced in Example \ref{RQO} of Section \ref{Algorithm design} is adopted for compressed communication. All numerical results are averaged over $S=50$ independent Monte Carlo trials to  eliminate random fluctuations.
We adopt the network-averaged mean-square error (MSE) as the performance metric:
$
\operatorname{MSE}(k) = \frac{1}{n}\sum_{i=1}^n \frac{1}{S}\sum_{s=1}^S \left\|\hat{\xi}_{k,i}^{(s)} - \xi_k\right\|^2,$
where $\hat{\xi}_{k,i}^{(s)}$ denotes the parameter estimate of sensor $i$ at time $k$ in the $s$-th Monte Carlo trial.

\begin{figure}[!htb]
	\centering
	\includegraphics[width=3.5in]{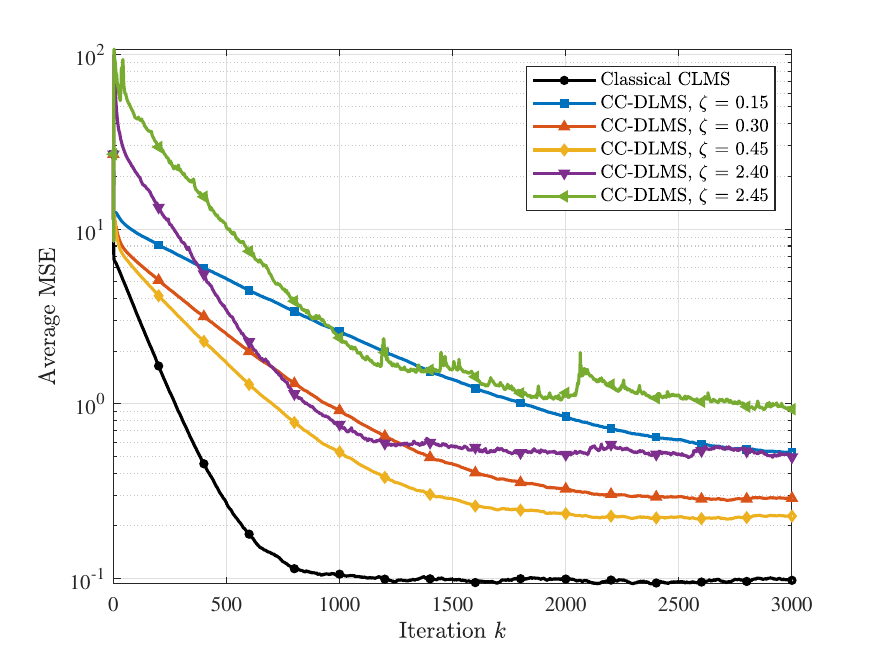}
	\caption{Network-averaged MSE for time-varying parameters ($\gamma=0.003$) with different compression weights $\zeta$.}\label{fig_zeta}
\end{figure}

1) \textbf{Effect of the compression weight \(\zeta\)}. We first examine the effect of the compression weight $\zeta$ in the slowly time-varying case with $\gamma=0.003$ and fixed quantization bit $r=2$. 
Five values \(\zeta \in \{0.15, 0.3, 0.45, 2.4, 2.45\}\) are tested. The set \(\zeta \in \{0.15, 0.3, 0.45\}\) satisfies the condition \(\zeta(1+\rho) < 2\), whereas \(\zeta \in \{2.4, 2.45\}\) violates it.
The classical consensus LMS (CLMS) algorithm without compression is included as a benchmark.
 Figure \ref{fig_zeta} plots the MSE trajectories of the classical CLMS and the proposed CC-DLMS with different $\zeta$. It can be observed that  all CC-DLMS  with \(\zeta \in \{0.15, 0.3, 0.45\}\) remain mean-square bounded, which corroborates the stability result in Theorem \ref{thm:consensus}. Within this stable regime, increasing \(\zeta\) accelerates the update of auxiliary compression variables, thus yielding faster convergence and lower tracking error.  By contrast, excessively large \(\zeta\) values (e.g., \(\zeta = 2.4\) and \(2.45\)) result in a  degradation of tracking performance. This phenomenon 
indicates that \(\zeta\) should be properly selected to balance auxiliary-state updating, compression-error correction, and tracking accuracy, as illustrated  in Remark \ref{remark6}.

\begin{figure}[!htb]
	\centering
	\includegraphics[width=3.5in]{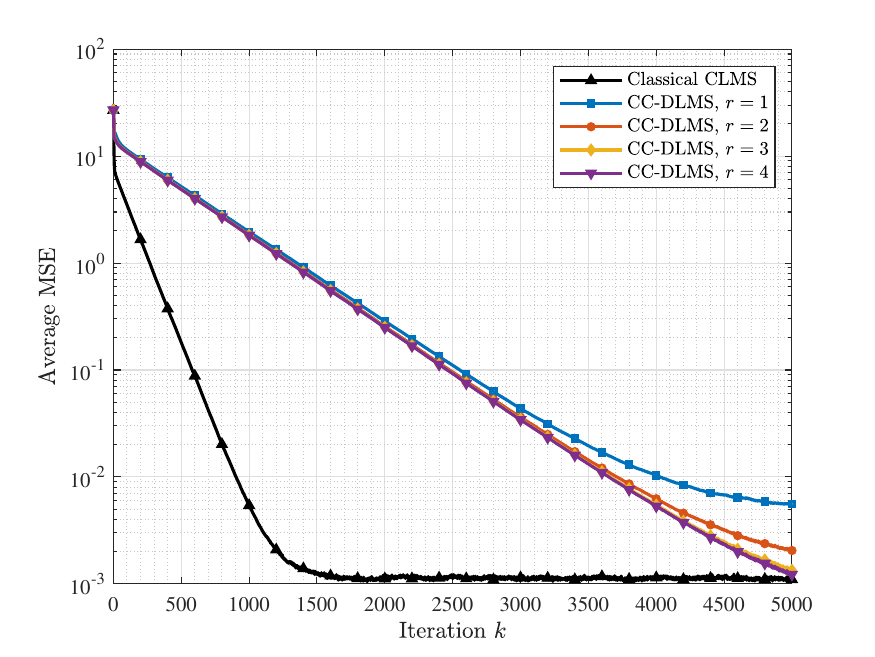}
	\caption{Network-averaged MSE for time-invariant parameters ($\gamma=0$) with different quantization bits $r$.}\label{fig_r0}
\end{figure}

\begin{figure}[!htb]
	\centering
	\includegraphics[width=3.5in]{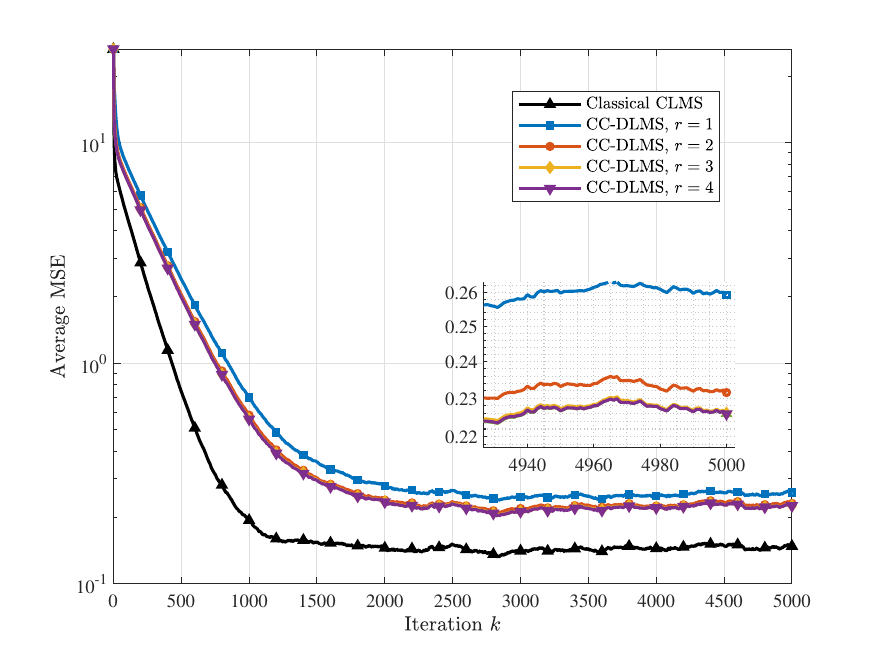}
	\caption{Network-averaged MSE for time-varying parameters ($\gamma=0.003$) with different quantization bits $r$.}\label{fig_r_gamma}
\end{figure}

2) \textbf{Effect of the quantization level \(r\).} We next investigate the trade-off between communication efficiency and estimation accuracy by varying the number of quantization bits $r \in \{1,2,3,4\}$, with a fixed compression weight $\zeta = 0.2$. Simulations are performed for both the time-invariant case ($\gamma=0$) and the time-varying case ($\gamma=0.003$), with results shown in  Figure \ref{fig_r0} and  Figure \ref{fig_r_gamma}, respectively. 
From  Figure \ref{fig_r0} and Figure \ref{fig_r_gamma}, the MSE decreases monotonically as $r$ increases.
For the time-varying case in  Figure \ref{fig_r_gamma}, the overall  MSE is higher than that in the time-invariant case, owing to the additional tracking error induced by parameter drift. 
Overall, these results illustrate the trade-off between communication cost and estimation accuracy controlled by the quantization level \(r\).

\section{Concluding remarks}\label{concluding}

This paper investigated distributed tracking of stochastic time-varying parameters under compressed communication.  
A differentially compressed LMS algorithm was developed using auxiliary compression states and randomized communication. 
Mean-square boundedness and explicit steady-state tracking-error bounds were established under cooperative excitation for connected undirected  networks, allowing correlated and nonstationary regressors without node-wise persistent excitation. 
The results  characterize the impacts of adaptation rate and compression parameters on  tracking performance.
Future work will exploit the derived tracking-error characterization to design adaptive compression and communication strategies.
Event-triggered and asynchronous mechanisms will also be explored for more resource-constrained networks.

\Acknowledgements{This work was supported by the  National Natural Science Foundation of China under Grants 62603361 and 62303452. }

\bibliographystyle{scis}
\begingroup
\makeatletter
\@for\revision@key:=Xie2022Distributed,Liu2026Distributed,gan2021stability,Liu2023Recursive,GAN2022105228\do{%
  \expandafter\let\csname revision@ref@\revision@key\endcsname\@empty}
\let\revision@bibitem\bibitem
\renewcommand{\bibitem}[2][]{%
  \normalcolor
  \ifcsname revision@ref@#2\endcsname\fi
  \if\relax\detokenize{#1}\relax
    \revision@bibitem{#2}%
  \else
    \revision@bibitem[#1]{#2}%
  \fi}
\makeatother
\bibliography{my_gd}
\endgroup

\end{document}